\documentstyle[11pt,paspconf,epsf]{article}
\def\ltsim{\raise 2pt \hbox {$<$} \kern-1.1em \lower 4pt \hbox {$\sim$}}
\def\gtsim{\raise 2pt \hbox {$>$} \kern-1.1em \lower 4pt \hbox {$\sim$}}
\def\edcomment#1{\iffalse\marginpar{\raggedright\sl#1\/}\else\relax\fi}
\reversemarginpar

\begin{document}
\title{VLBI Observations of Mkn 501 and Mkn 421}
\author{G. Giovannini, L. Feretti and T. Venturi}
\affil{Istituto di Radioastronomia, via Gobetti 101, 40129 Bologna, Italy}
\author{W.D. Cotton}
\affil{NRAO, 520 Edgemont Rd, Charlottesville VA 22903-2475, USA}
\author{L. Lara}
\affil{IAA, CSIC, Apdo. 3004, 18080 Granada, Spain}

\begin{abstract}
We present here two epochs of Space VLBI Observations at 18 cm of the BL-Lac
type object Mkn 501. Thanks to the high resolution of these new data we have
found that the inner jet is centrally brightened at its beginning but becomes
extended and limb brightened at $\sim$ 8 mas from the core. Moreover a 
comparison
between the two epochs shows the presence of a possible 
proper motion with apparent
velocity = 6.7c. 
VLBI data at 6 and 18 cm of the BL-Lac
type object Mkn 421 are also presented.
Observational data have been used to constrain the jet
velocity and orientation.
\end{abstract}

\section{Introduction and Observations}

Mkn 501 and Mkn 421 are two nearby BL-Lac 
type objects at z = 0.034 and 0.030, respectively. They are well
studied sources in radio, optical and X-ray bands being among the brightest
Bl-Lac objects at all wavelengths. Mkn 421 was the first source
to be detected at TeV energies, followed by Mkn 501.
In the radio band, both sources show a kpc scale morphology and a total radio 
power consistent with the expectation of unified scheme models that BL-Lac type
sources are FR I galaxies oriented at small angles to the line of sight.

We present here new Space VLBI observations of Mkn 501 and VLBI data of 
Mkn 421. Mkn 501 was observed at 18 cm on August 4th, 1997 and April 8th, 1998
in a space VLBI project using the HALCA satellite and 12 ground
stations.
Mkn 421 was observed at 6 cm in July 1995 with the VLBA and at 18cm in February
1996 with the global VLBI.
The data of both sources were correlated in Socorro (NM - USA) and reduced
with the AIPS package.

A Hubble costant of 50 km sec$^{-1}$ Mpc$^{-1}$ is assumed throughout. 

\section{Jet Morphology}

{\it Mkn 501}. 
At parsec resolution Mkn 501 shows a one-sided jet  which, at $\sim$ 20 mas from the
core, changes
its orientation from $\sim$ 140$^\circ$ to $\sim$ 30$^\circ$.
\begin{figure}
\plottwo{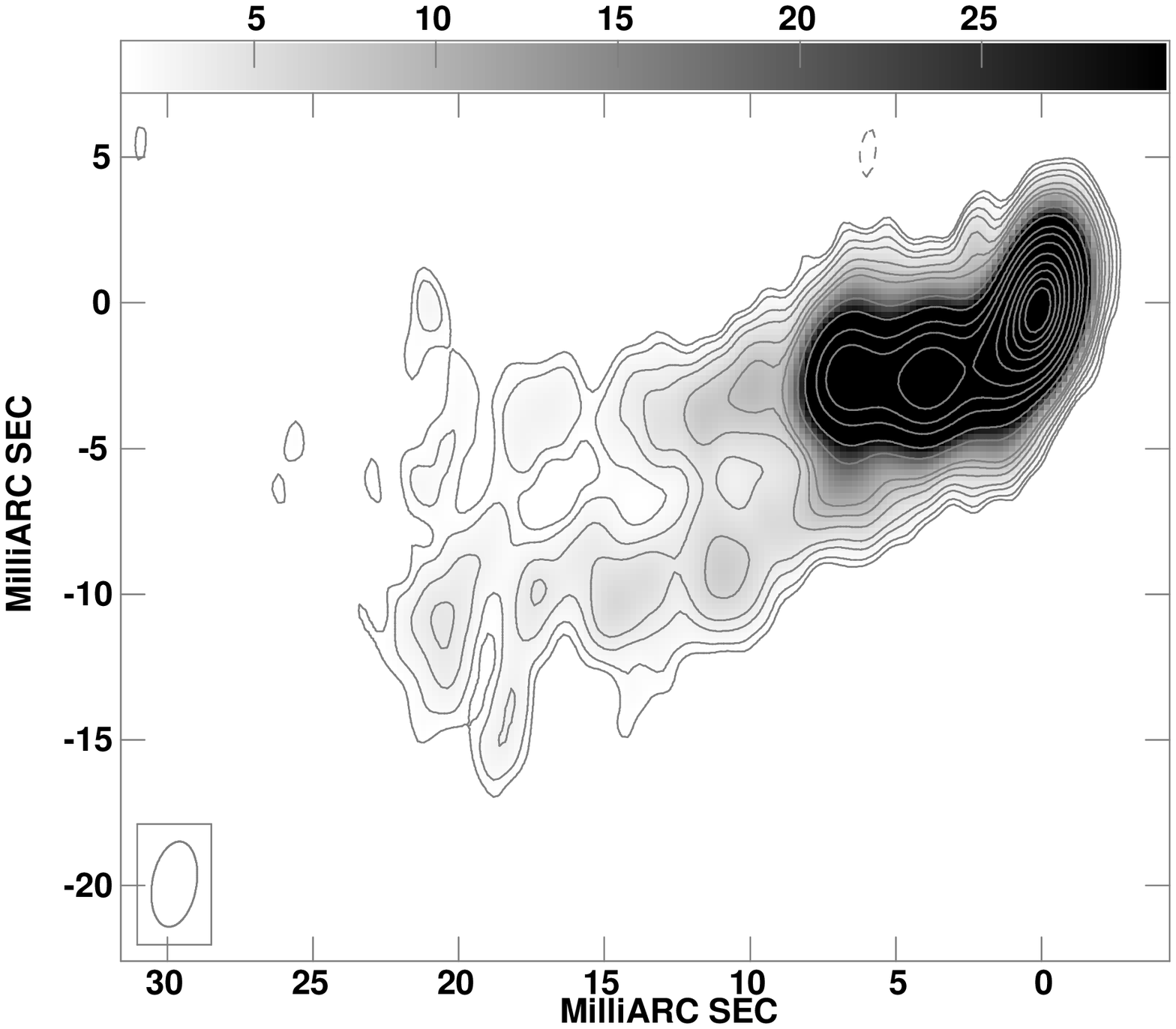}{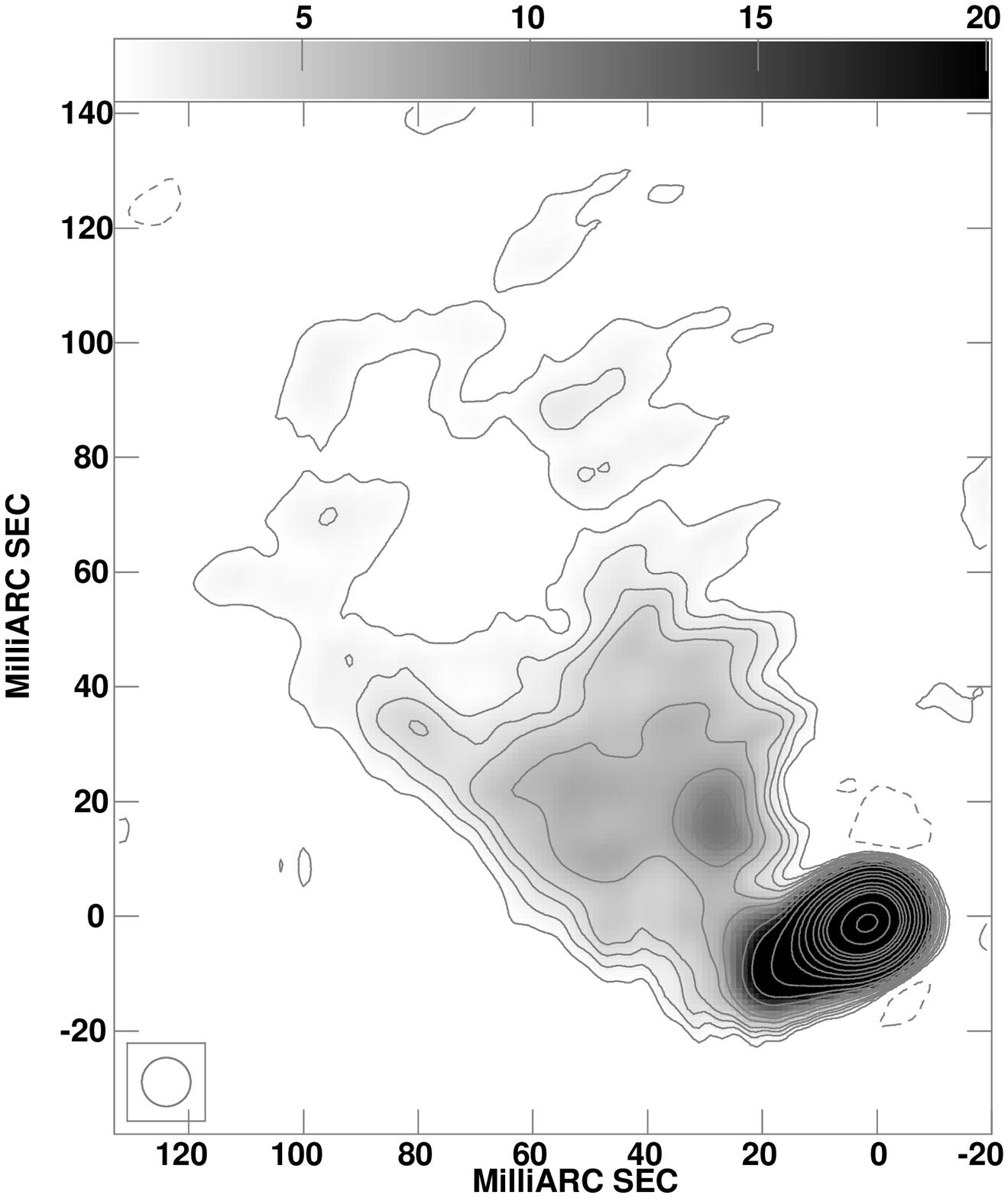}
\caption{ 
Left: Space VLBI map of Mkn 501 at 18cm. The HPBW is 2.9 $\times$
1.5 mas at PA = -10 $^\circ$. The noise level is 0.3 mJy/beam. 
Right: VLBI 
map of Mkn 501 at 18 cm using only ground stations. The HPBW is
8.5 mas and the noise level is 0.35 mJy/beam.
}
\end{figure}
From the analysis of the map (Fig. 1-left) we note a clear 
change in the jet structure: at the beginning
the jet is resolved and its brightness is centrally peaked, but at $\sim$ 8
mas from the core, it 
becomes limb-brightened with the maximum of
the surface brightness on both sides of the parsec scale jet.
We interpret this observational result as indication of a change
in the jet physical properties at $\sim$ 8 mas (7.4 pc) 
from the core. 
The limb-brightened jet is well visible for 20-25 mas ($\sim$ 20 pc).
This result is in agreement with the polarization properties of this source
presented by Aaron et al. (present proceedings).  
At a larger distance from the core ($\sim$ 20 pc), the jet shows a large 
change in its position 
angle and a dramatic expansion. In our low resolution map (Fig. 1-right) 
the jet is visible for more than 
100 mas and appears still edge-brightened. No evidence of a helical
structure is visible in our maps.
\goodbreak
\noindent
{\it Mkn 421}. At 6 cm Mkn 421 shows a core emission and a one-sided jet.
The jet is well collimated at the beginning, but 
at $\sim$ 5 mas (4 pc) from the core it shows many wiggles and changes in
the position angle (Fig. 2-left). 
This complex structure is confirmed by the 18 cm map 
(Fig. 2-right) at lower resolution and on a larger scale. 
At $\sim$ 20 mas from the 
core, the parsec scale jet shows a dramatic expansion, its transversal
size being comparable with its longitudinal dimension. 
Present data do not have enough resolution or sensitivity to distinguish 
between a centrally peaked or a limb-brightened jet.
\begin{figure}
\plottwo{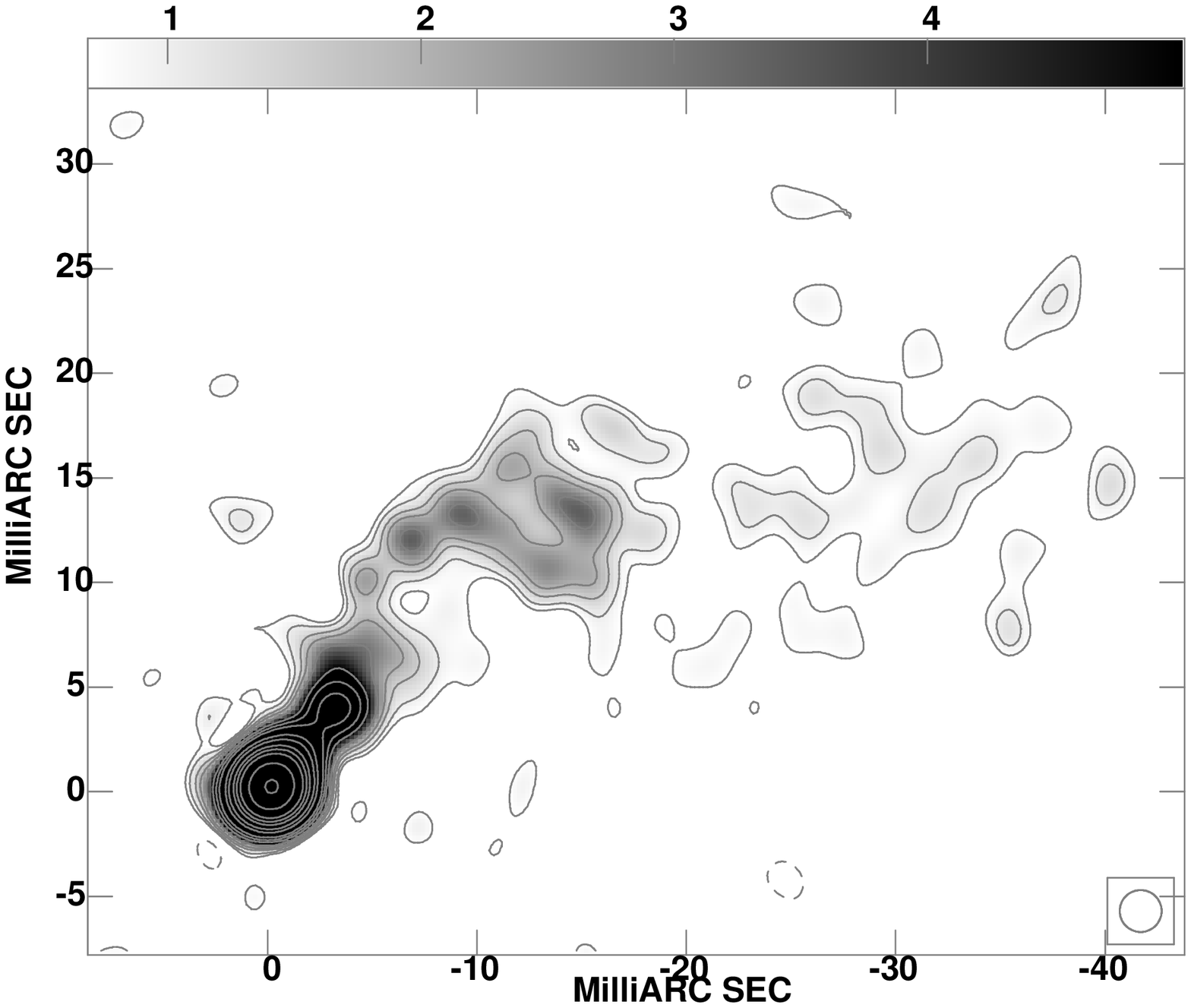}{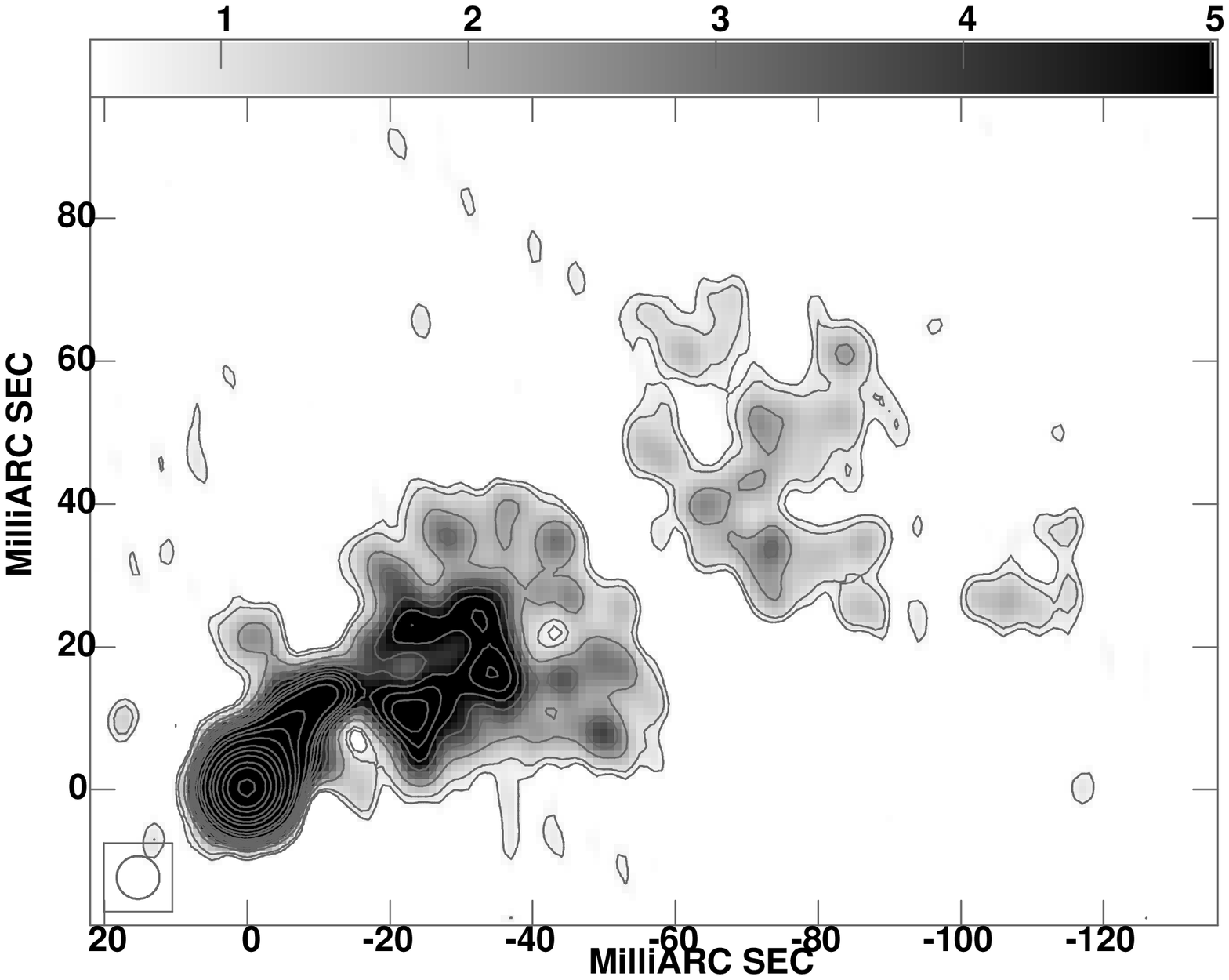}
\caption{ 
Left: VLBA map of Mkn 421 at 6 cm. The HPBW is 2.0 mas; the noise level is 
0.24 mJy/beam. 
Right: Global VLBI map at 18 cm. The HPBW is 6 mas; the noise level is 0.3
mJy/beam. 
}
\end{figure}
\section{Jet Orientation and Velocity}

We used the available VLBI data to derive information on the jet velocity and
orientation.
Assuming that parsec scale jets are intrinsically two-sided and symmetric, we
can constrain the jet velocity ($\beta$ = v/c) and inclination to the
line of sight ($\theta$) from the observed jet asymmetry.
For {\bf Mkn 501} the Jet/CounterJet ratio is \gtsim 200, which leads to 
$\beta$ cos$\theta$ \gtsim 0.79. Moreover, from the known 
correlation between the core power and the unbeamed total
radio power (Giovannini et al., 1988) we can infer
the expected intrinsic core radio power from the observed total radio power. 
Comparing the expected and the measured core radio power,
we obtain that Mkn 501 has to be oriented at $\theta$ \ltsim 26$^\circ$ with 
$\beta$ in the range 1 - 0.88.
We produced identical maps from the first and second epoch space VLBI data of
Mkn 501 at 18 cm. A comparison of
the two maps suggests the existence of a possible proper motion between the 
two epochs with apparent velocity $\sim$ 6.7c. We are 
aware that to measure
a proper motion with only two epoch data may give unreliable results, however
we note that the present data are of good quality and the structures in the 
two maps are in good agreement.
More Space VLBI Observations have been requested to confirm this result.
Such an apparent velocity implies that the real jet velocity has 
to be \gtsim 0.989c and the jet has to be oriented at an angle smaller than
17$^\circ$.
If we assume that the bulk and pattern jet velocity are comparable,
we can compare the observational constraints previously
discussed with the measured proper motion. From these results we derive that 
Mkn 501 is oriented at $\sim$ 10$^\circ$ - 15$^\circ$ with a velocity in the 
range
0.990 - 0.999c. It implies high values of the Lorentz factor: $\gamma \sim$
7 - 22, but a relatively low Doppler factor: $\delta \sim$ 1.3 - 5.6.
From the jet sidness and core dominance of {\bf Mkn 421}
we derive for this source an inclination
angle $\theta$ $\ltsim$ 30$^\circ$
and $\beta$ in the range 1 - 0.84.
The measured proper
motion for Mkn 501 and the derived intrinsic jet velocities suggest that Bl-Lac type 
objects may have parsec scale jets with intrinsic velocities of the same order of 
parsec scale jets in radio quasars. 

\section{Discussion and Conclusions}

Thanks to the high resolution provided by the VSOP we have imaged 
the parsec scale jet of Mkn 501 in great detail.
It shows a change in its physical properties at 8 mas from the
core: the innermost jet is centrally brightened, but at $\sim$ 8 mas from the 
core it becomes limb-brightened. 
Such a structure is expected by the {\it central spine -- shear layer} 
jet model
(Laing, 1996). The observational data discussed here imply that the shear layer
is not visible at the jet beginning but appears at $\sim$ 7 pc 
from the core 
corresponding to a deprojected distance of 40 - 45 pc assuming $\theta$ = 
10$^\circ$. 
At $\sim$ 20-25 mas from the core, the jet shows a prominent bend and a rapid
expansion. 

A similar expansion is found also in Mkn 421, where the jet 
dramatically
expands at about the same distance from the core.
We interpret this result as evidence of 
a strong interaction with the surrounding medium which affects the jet 
dynamics, causing the jet velocity decrease observed in FR I radio galaxies. 

Current models of gamma ray emission and intra-day 
variability suggest for these sources that in the innermost region
(0.003 - 0.03 pc) the jet should be oriented at $\theta$ \ltsim 5.7 $^\circ$
and should have a Lorentz factor $\gamma$ \gtsim 10 which implies $\beta \sim$
0.995 and $\delta \sim$ 10 (see Spada et al. present proceedings and Salvati 
et al., 1998). These results seem to be in contrast with the derived 
constraints from VLBI radio data, however we note that the regions
where the radio emission ({\it the core region}) and the gamma and X-Ray emission
are produced, are not coincident, the former being at a larger distance 
from the active core than the latter.
We estimate that the {\it radio core} has a size \gtsim 0.05 pc. 
Therefore radio and gamma ray data are in agreement if when travelling between
the distance of 0.003 pc to 0.05 pc, the jet orientation changes from 
$\sim$ 5$^\circ$ to $\sim$
10$^\circ$. Such a change in the jet orientation is plausible
given the distorted
morphology found on the parsec scale of Mkn 501 and Mkn 421. 

\acknowledgements  
We acknowledge partial financial support from Italian MURST and ASI. 
We gratefully acknowledge the VSOP Project, which is led by the Japanese
Institute of Space and Astronautical Science in cooperation with many
organizations and radio telescopes around the world. The NRAO is operated by
Associated Universities, Inc. under a cooperative agreement with the National
Science Foundation.

\end{document}